\documentclass[useAMS,usenatbib]{mn2e}
\usepackage{times}

\input amssym.tex

\usepackage{graphics}
\usepackage{graphicx}
\usepackage{epsf}
\usepackage{url}

\newcommand{\bk}{{\bmath k}}
\newcommand{\ba}{{\bmath a}}
\newcommand{\bb}{{\bmath b}}
\def\LCDM{$\Lambda$CDM}
\def\hMpc{{\ }h^{-1}{\,}{\rm Mpc}}
\def\ihMpc{{\ }h{\,}{\rm Mpc}^{-1}}
\def\nospaceihMpc{h{\,}{\rm Mpc}^{-1}}
\def\hMpcV{{\ }h^{-3}{\,}{\rm Mpc}^3}
\def\ihMpcV{{\ }h{\,}{\rm Mpc}^{-3}}
\def\hGpc{{\ }h^{-1}{\,}{\rm Gpc}}
\def\hGpcV{{\ }h^{-3}{\,}{\rm Gpc}^3}

\def\bssC{\textbf{\textsf{C}}}
\def\R{\mathcal{R}}
\def\halofit{{\scshape halofit}}
\def\camb{{\scshape camb}}
\def\ug{u_{\rm g}}

\newcommand{\ptcovarwig}{
  \begin{figure}
    \begin{center}
     \leavevmode
      \epsfxsize=\columnwidth
      \epsfbox{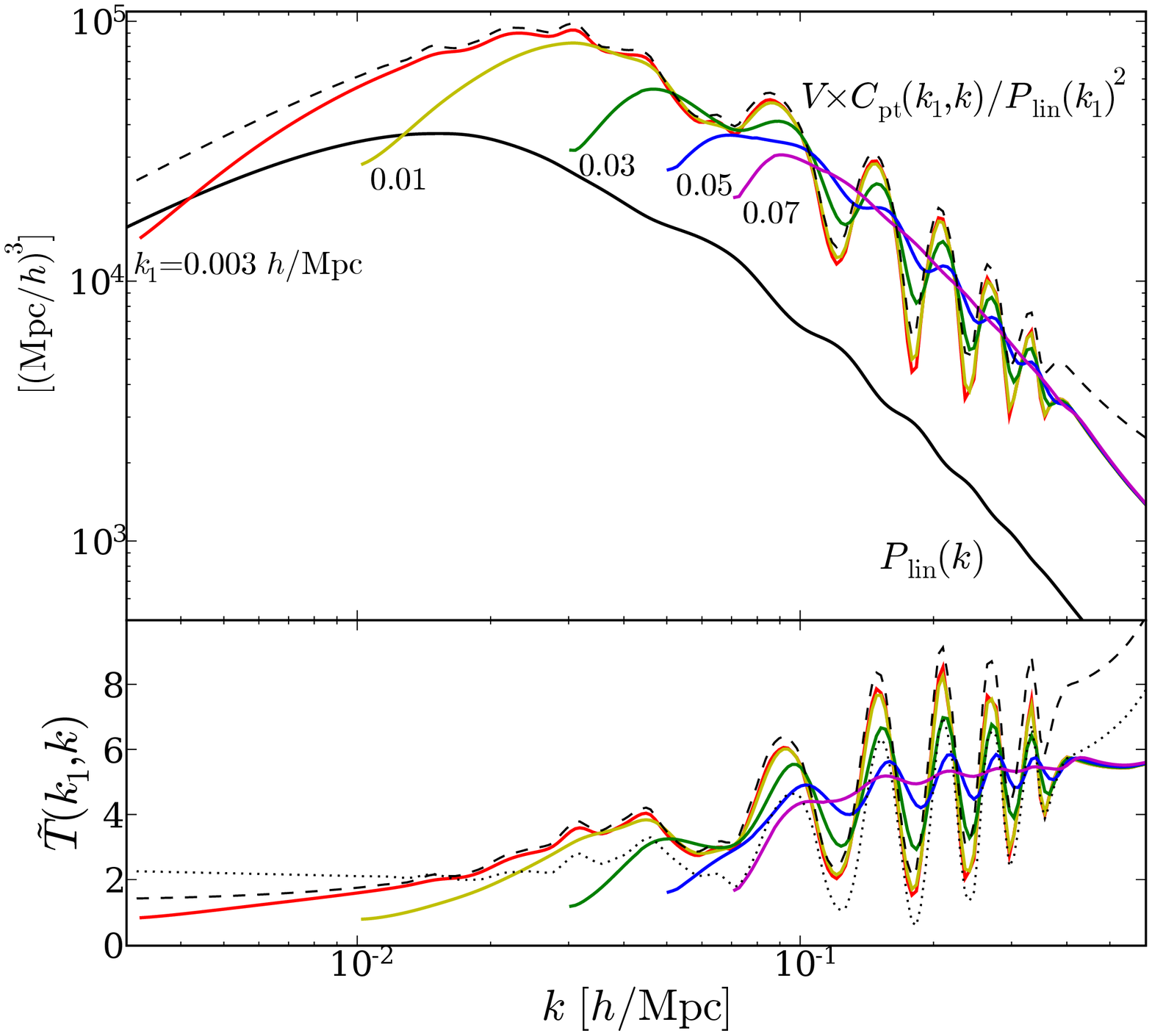}
    \end{center}

    \caption[1]{ \small Baryon acoustic oscillations in off-diagonal
      terms of the power-spectrum covariance matrix, calculated using
      leading-order perturbation theory for the trispectrum.  In the
      top panel, each colored curve is a row of the covariance matrix,
      divided by $P_{\rm lin}(k_1)^2$, where $k_1$ is the wavenumber
      of that row.  In the bottom panel, the curves are additionally
      divided by $P_{\rm lin}(k)$.  The dashed curves show the
      approximation of Eq.\ (\ref{tapprox}).  The dotted curve in the
      bottom panel shows the approximation without the
      first-derivative term.
    \label{ptcovarwig}
    }
  \end{figure}
}
\newcommand{\hmcovarwig}{
  \begin{figure}
    \begin{center}
     \leavevmode
      \epsfxsize=\columnwidth
      \epsfbox{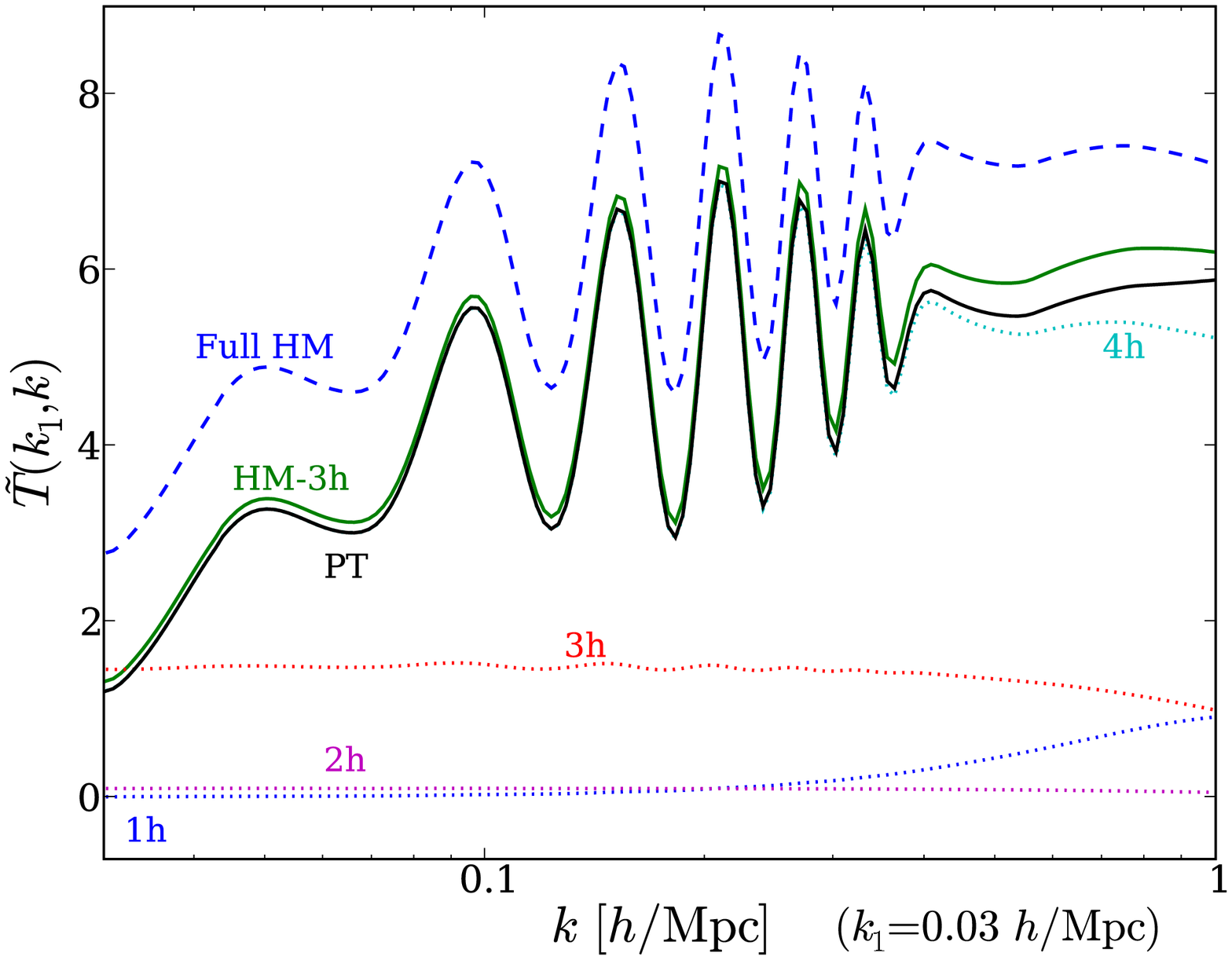}
    \end{center}

    \caption[1]{ \small A comparison of different terms in the
      halo-model prediction for the normalized power-spectrum
      covariance $\tilde{T}(k_1,k)=C(k_1,k)V/[P_{\rm lin}(k_1)^2
      P_{\rm lin}(k)]$ at $z=0$.  We use $P_{\rm lin}$ instead of the
      halo-model matter $P$ in the denominator to show clearly how
      $C(k_1,k)$ differs from the perturbation-theory prediction.  On
      large scales, the 1h and 2h terms are negligible, but the 3h
      term gives a probably inaccurate additional contribution.  On
      smaller scales (both larger $k_1$ and $k$) than those shown,
      the 1h term comes to dominate $C(k_1,k)$.
    \label{hmcovarwig}
    }
  \end{figure}
}
\newcommand{\hmgalcovarwig}{
  \begin{figure}
    \begin{center}
     \leavevmode
      \epsfxsize=\columnwidth
      \epsfbox{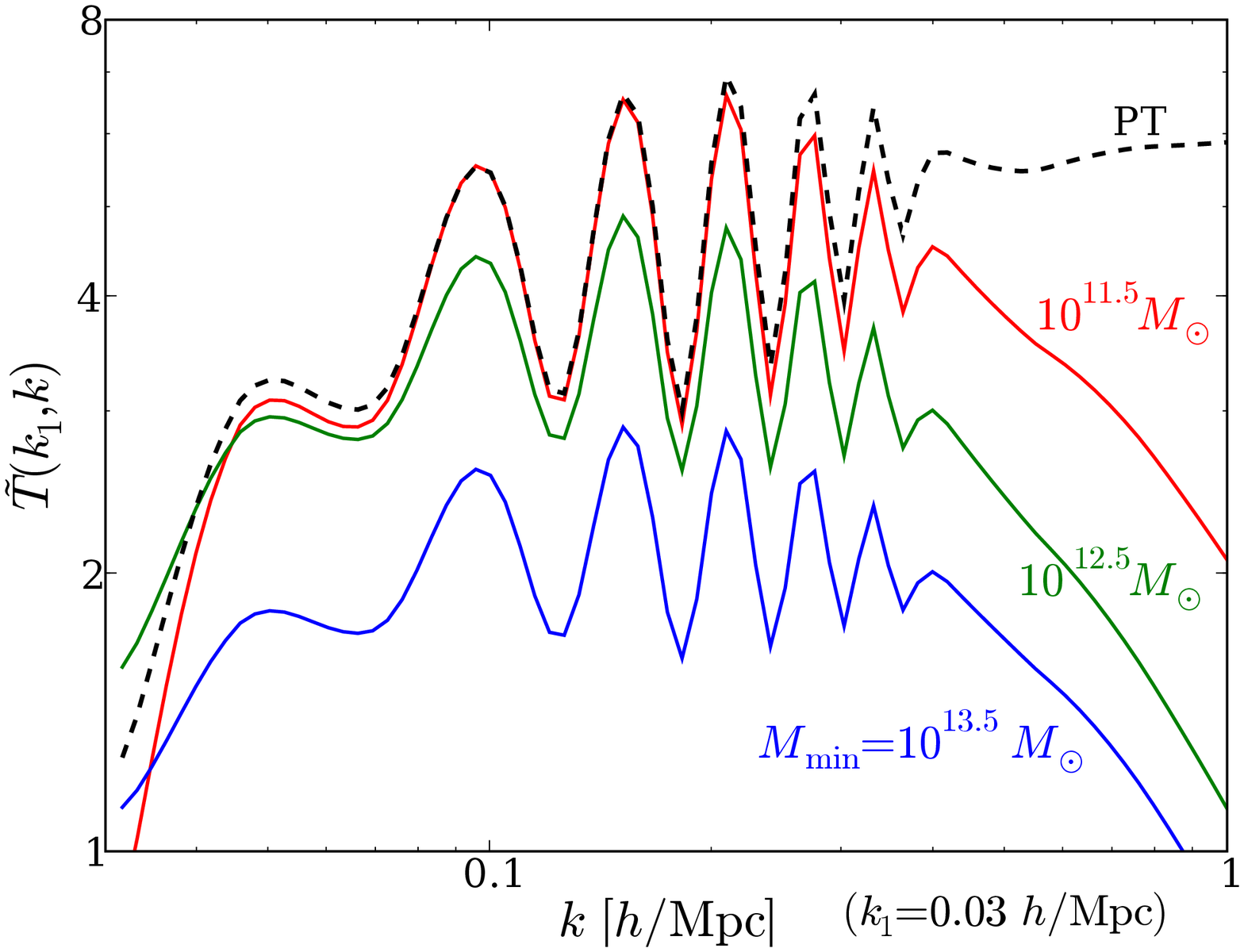}
    \end{center}

    \caption[1]{ \small Wiggles in off-diagonal elements of the
    covariance matrix of halo-model galaxy power spectra, as shown for
    the matter power spectrum in Fig.\ \ref{hmcovarwig}.  A few
    different Halo Occupation Distributions are shown, labeled by the
    minimum mass for a halo to host a central galaxy.
    \label{hmgalcovarwig}
    }
  \end{figure}
}
\newcommand{\showpows}{
  \begin{figure}
    \begin{center}
     \leavevmode
      \epsfxsize=\columnwidth
      \epsfbox{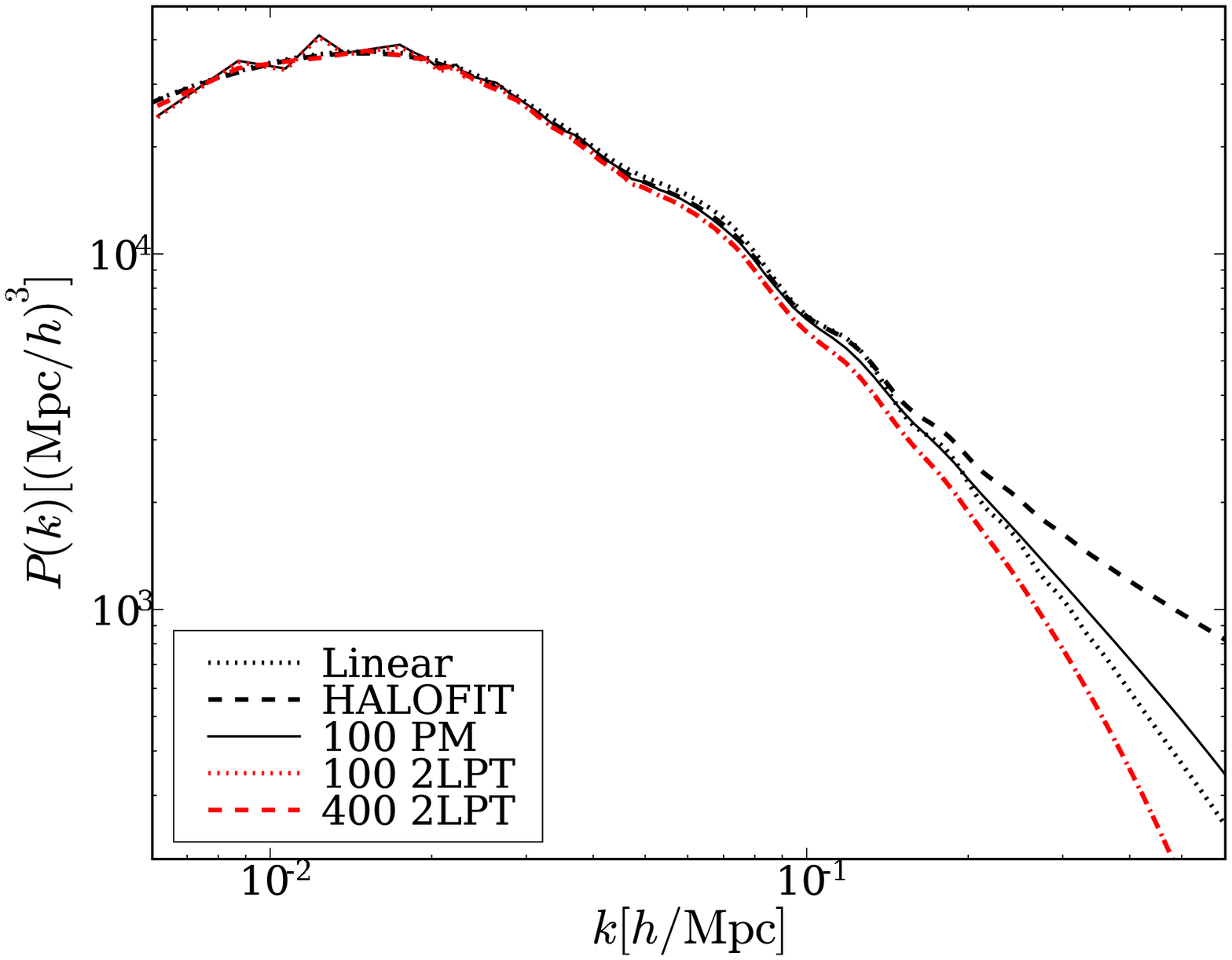}
    \end{center}

    \caption[1]{ \small Power spectra of simulations used in the
    paper, averaged over the 100 or 400 realizations in each ensemble.
    The pure-2LPT simulations begin a gradual attenuation at rather
    large scales ($k\approx 0.05\ihMpc$), while the particle mesh (PM)
    simulations seem trustworthy to $k\approx 0.14\ihMpc$.
    \label{showpows}
    }
  \end{figure}
}
\newcommand{\bigbins}{
  \begin{figure}
    \begin{center}
     \leavevmode
      \epsfxsize=\columnwidth
      \epsfbox{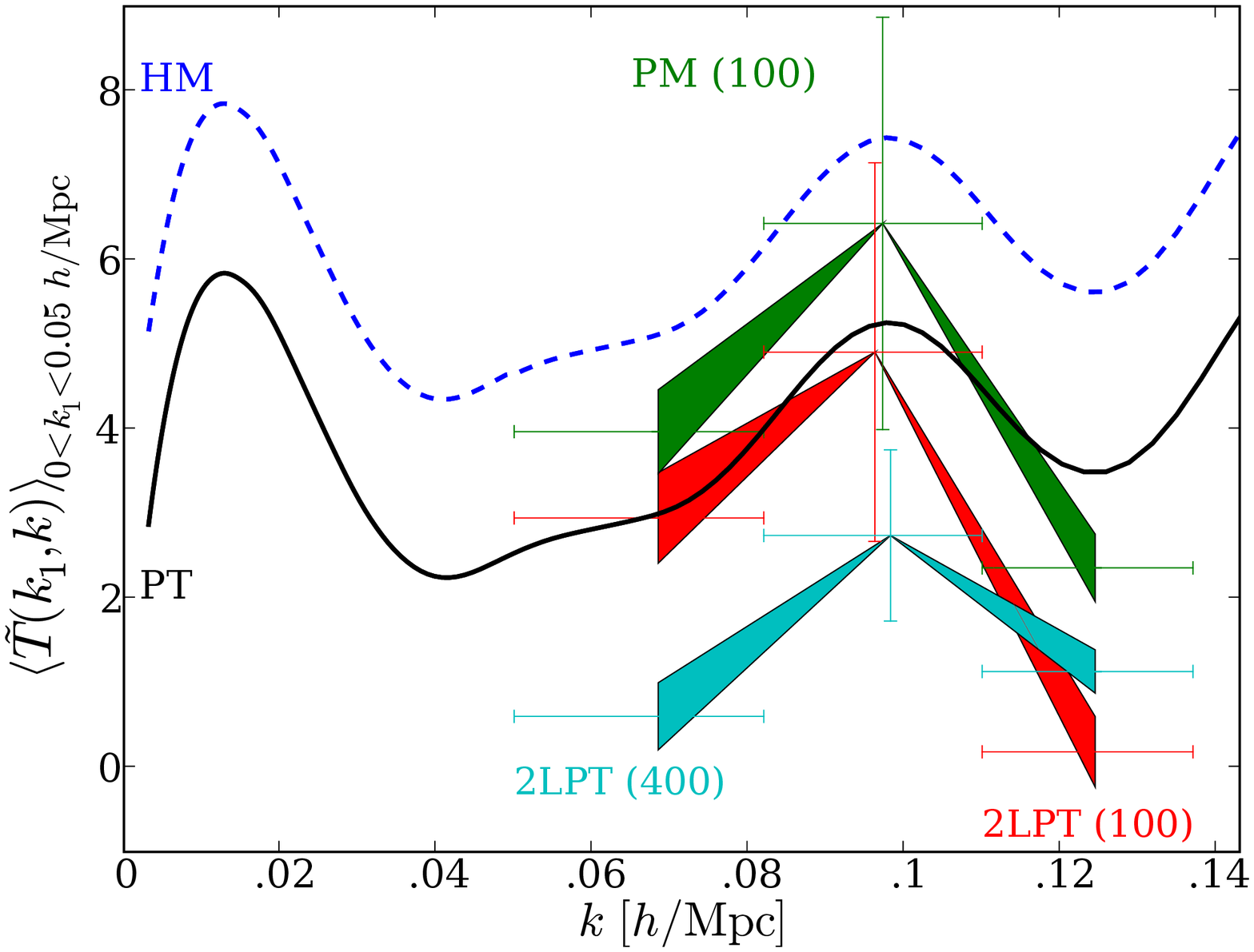}
    \end{center}

    \caption[1]{ \small Measurements of the normalized power-spectrum
    covariance $\tilde{T}$ from ensembles of $N$-body simulations.
    $\tilde{T}$ is measured in large bins to beat down noise, with
    edges roughly where nodes in $\tilde{T}$'s wiggles are expected.
    Only the middle points have error bars, which are typical of all
    three points, since different jackknife samples typically move up
    and down together.  The cones coming from the middle points
    indicate the degree of robustness in the peak; see text for
    details.  From 100 PM simulations of box size $1024\hGpc$, we estimate a detection of the
    peak at a 5-sigma level.  The simulations evolved entirely using
    2LPT have somewhat lower values of $\tilde{T}$; the 100 2LPT
    simulations and the 100 PM simulations had the same initial
    conditions.  With 300 additional 2LPT simulations, the noise in
    $\tilde{T}$ is reduced, and the peak shape is roughly that of the
    PT prediction.
    
    \label{bigbins}
    }
  \end{figure}
}
\newcommand{\fineweights}{
  \begin{figure}
    \begin{center}
     \leavevmode
      \epsfxsize=\columnwidth
      \epsfbox{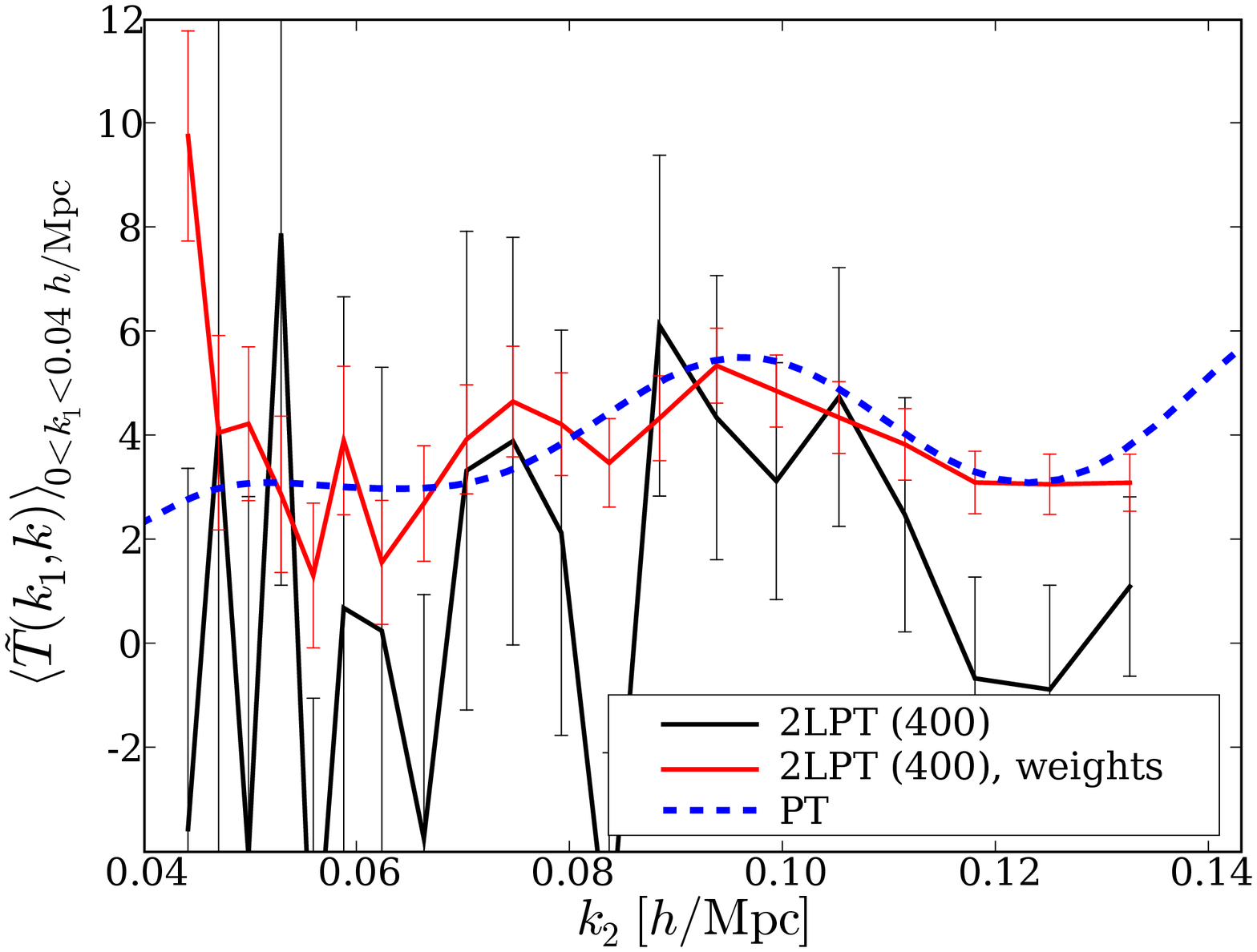}
    \end{center}

    \caption[1]{ \small Higher-resolution (more-finely binned)
    measurements of the normalized power-spectrum covariance
    $\tilde{T}$ from 400 2LPT simulations.  The red curve is
    $\tilde{T}$ measured using 52 weightings per simulation, according
    to the prescription of \citet{hrs}.  The red curve is probably
    raised from the true level of $\tilde{T}$ in the 2LPT simulation
    because of the beat-coupling which the weightings induce.
    \label{fineweights}
    }
  \end{figure}
}
\newcommand{\fbinfo}{
  \begin{figure}
    \begin{center}
     \leavevmode
      \epsfxsize=\columnwidth
      \epsfbox{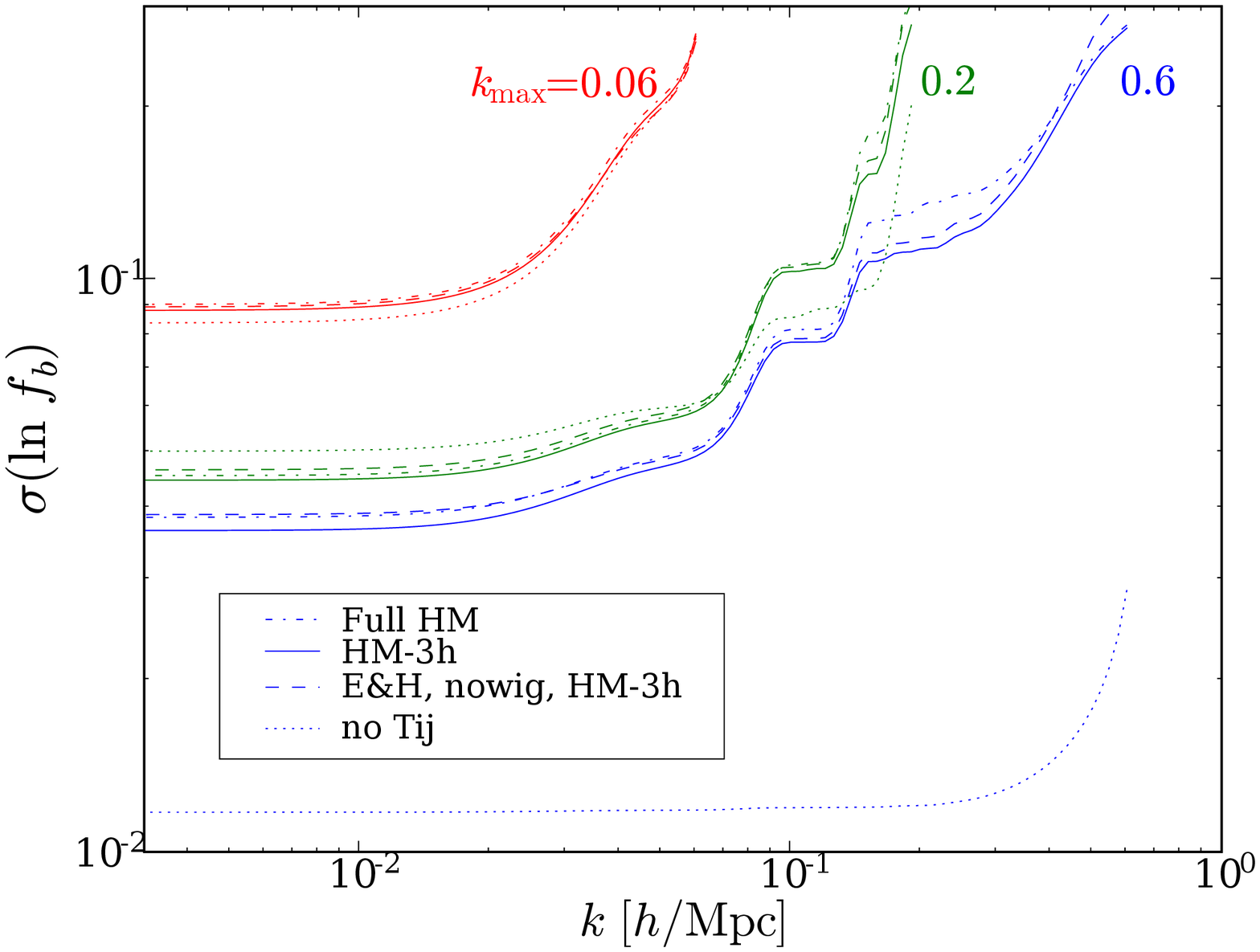}
    \end{center}

    \caption[1]{ \small The effect of using different covariance
      matrices when analyzing matter power spectra to constrain the
      logarithm of the baryon fraction.  Each curve starts at a
      $k_{\rm max}$, and shows how error bars tighten as larger scales
      are included in the analysis.  The fiducial covariance matrix we
      recommend (the halo model covariance, minus the 3h term) has
      been used for the solid curves.  A no-wiggle power spectrum was
      used to produce the covariance matrices used for the dashed
      curves.  Only the Gaussian variance is used for the dotted
      curves.  The wiggles in the covariance matrix reduce $\sigma(\ln
      f_b)$ by 7\% for $k_{\rm max}=0.6$.
    \label{fbinfo}
    }
  \end{figure}
}
\newcommand{\shinfo}{
  \begin{figure}
    \begin{center}
     \leavevmode
      \epsfxsize=\columnwidth
      \epsfbox{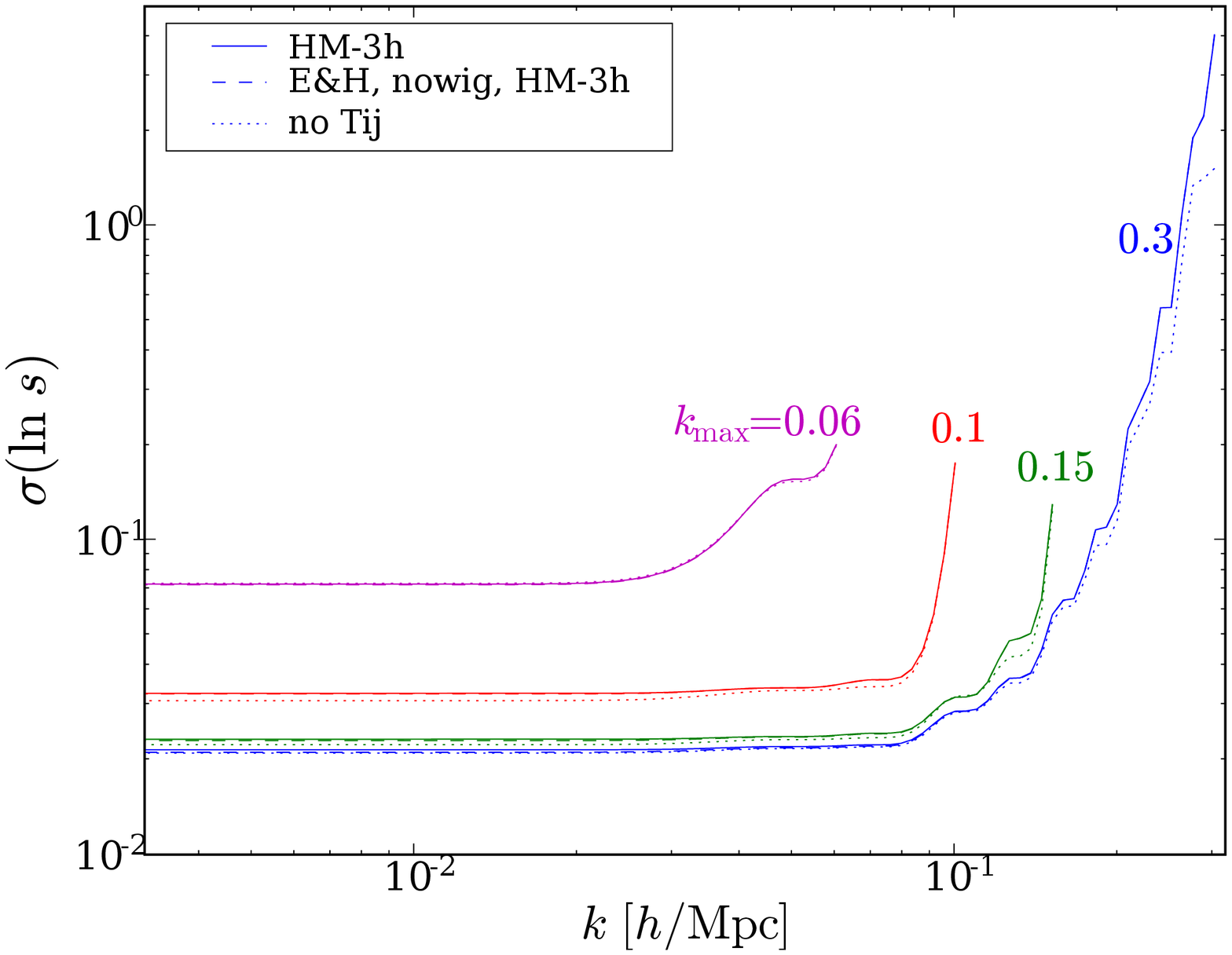}
    \end{center}
    \caption[1]{ \small As in Fig.\ \ref{fbinfo}, except showing error
    bars on the baryon acoustic scale (actually, the logarithm of the
    sound-horizon scale).  Here, the non-Gaussian covariance matrix
    terms are essentially negligible.  The wiggles in the covariance
    matrix enlarge $\sigma(\ln f_b)$ by 2\% for $k_{\rm max}=0.3$.
    \label{shinfo}
    }
  \end{figure}
}
\newcommand{\hods}{
  \begin{figure}
    \begin{center}
      \leavevmode
      \epsfxsize=\columnwidth
      \epsfbox{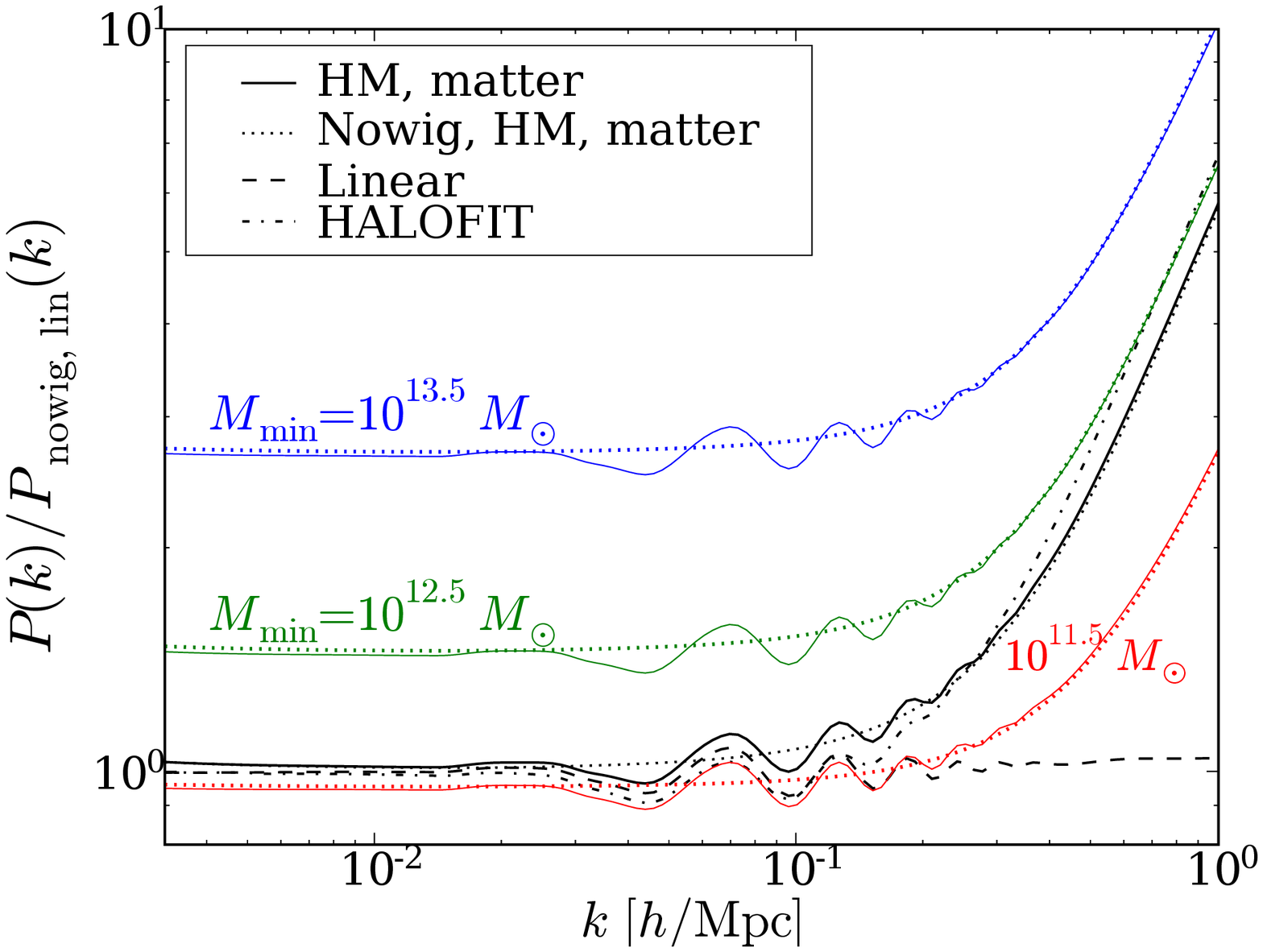}
    \end{center}
    \caption[1]{ \small Various halo-model power spectra used in the
      paper, divided by an \citet{ehu} no-wiggle power spectrum.  The
      colored curves are halo-model galaxy power spectra, labeled by
      the minimum mass of a halo which hosts a central galaxy.
      \label{hods}
    }
  \end{figure}
}
\newcommand{\galfbinfo}{
  \begin{figure}
    \begin{center}
     \leavevmode
      \epsfxsize=\columnwidth
      \epsfbox{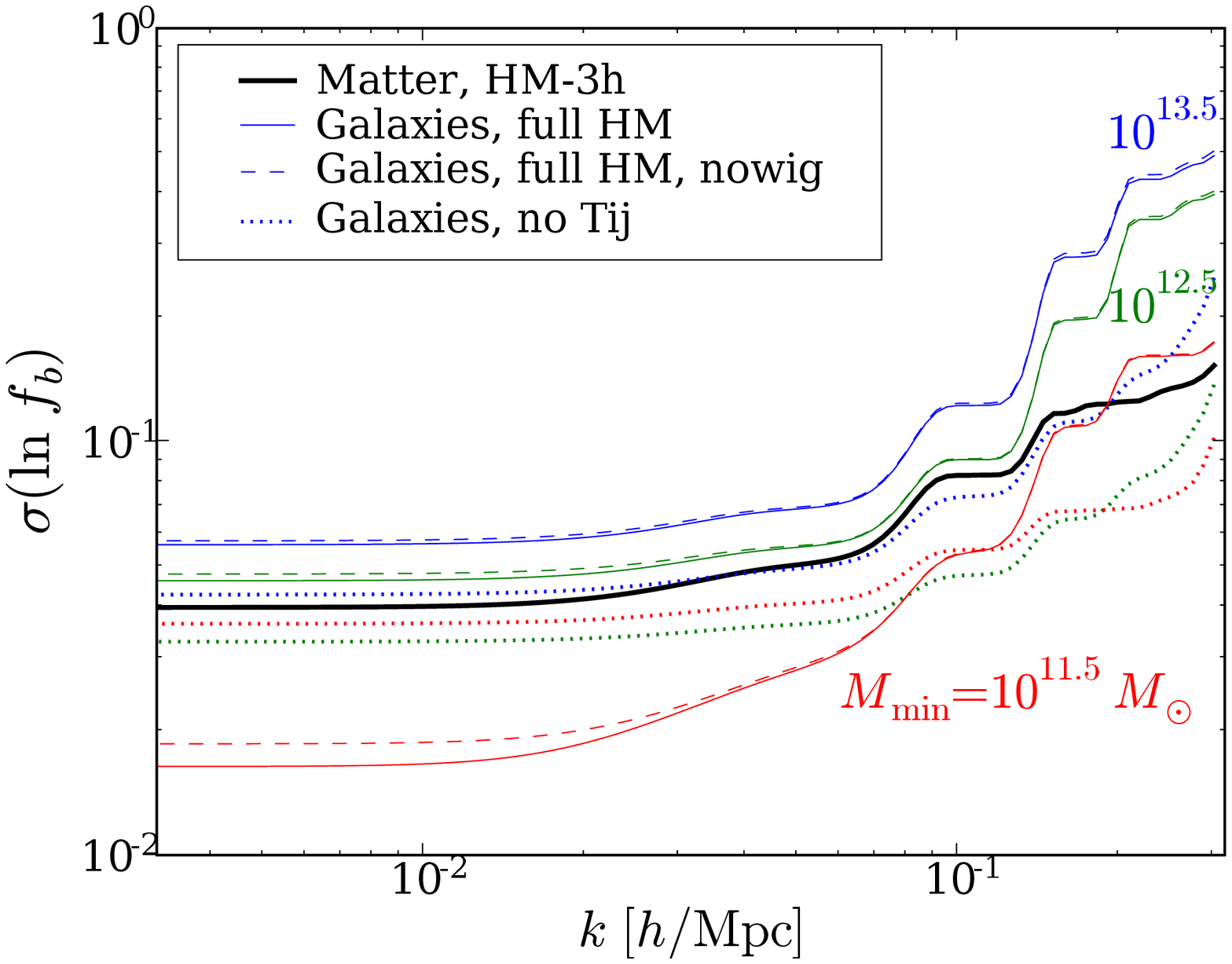}
    \end{center}

    \caption[1]{ \small At fixed volume, error-bar widths on the logarithm
    of the baryon fraction $f_b$ from analyzing power spectra of a few
    different galaxy samples.  This is as in Fig.\ \ref{fbinfo}, which
    shows error-bars for matter power spectra, except here, $k_{\rm
    max}$ is fixed at $0.3\ihMpc$.  The wiggles in the covariance
    matrix for galaxies have about the same influence on the
    error-bars on $\ln f_b$ as for matter, if not a bit more.  Also, a
    significant reduction in error-bar width occurs for low $M_{\rm
    min}$ (the minimum halo mass which can support a central galaxy).
    \label{galfbinfo}
    }
  \end{figure}
}
\newcommand{\galshinfo}{
  \begin{figure}
    \begin{center}
     \leavevmode
      \epsfxsize=\columnwidth
      \epsfbox{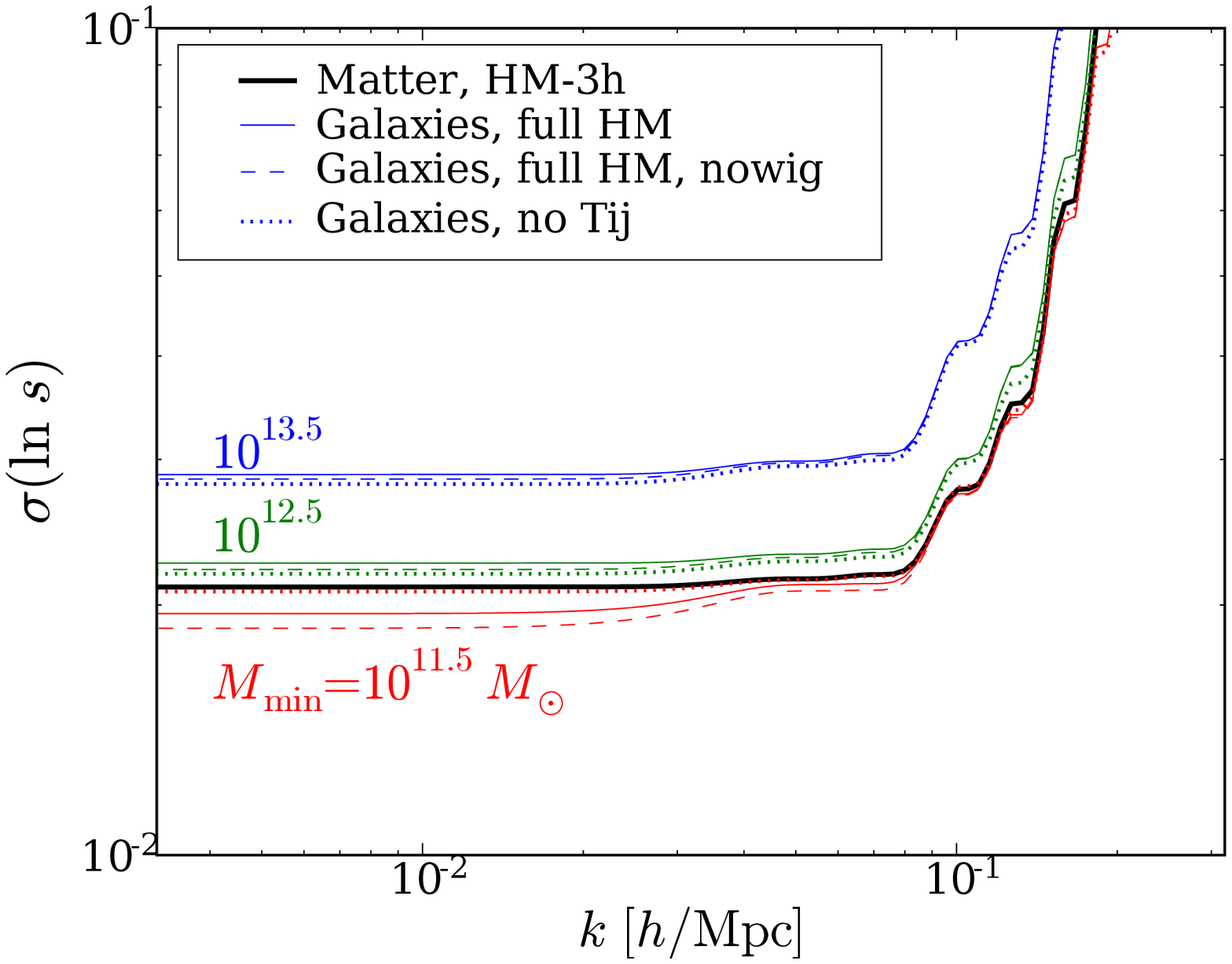}
    \end{center}

    \caption[1]{ \small At fixed volume, error-bar widths on the
    logarithm of the sound horizon $s$ from analyzing power spectra of
    a few different galaxy samples.  The wiggly covariance matrix here
    has a bit greater effect for galaxies than for matter; for
    comparison, see Fig. \ref{shinfo}, but notice that the $y$-axis is
    inflated here relative to that figure.
    \label{galshinfo}
    }
  \end{figure}
}

\begin{document}
\title[Baryon oscillations in power-spectrum covariance] {Baryon oscillations in galaxy and matter power-spectrum covariance matrices}

\author[Mark C.\ Neyrinck and Istv\'{a}n Szapudi]
{Mark C.\ Neyrinck$^1$ and Istv\'{a}n Szapudi$^1$\\
$^{1}$Institute for Astronomy, University of Hawaii, Honolulu, HI 96822, USA\\
  {\small \rm email: {\tt neyrinck@ifa.hawaii.edu}}. \rm The definitive version is available at {\tt www.blackwell-synergy.com}.}

\pubyear{2008}

\bibliographystyle{mnras}
	 
\maketitle
  
\begin{abstract}
We investigate large-amplitude baryon acoustic oscillations (BAO's) in
off-diagonal entries of cosmological power-spectrum covariance
matrices.  These covariance-matrix BAO's describe the increased
attenuation of power-spectrum BAO's caused by upward fluctuations in
large-scale power.  We derive an analytic approximation to
covariance-matrix entries in the BAO regime, and check the analytical
predictions using $N$-body simulations.  These BAO's look much
stronger than the BAO's in the power spectrum, but seem detectable
only at about a one-sigma level in gigaparsec-scale galaxy surveys.
In estimating cosmological parameters using matter or galaxy power
spectra, including the covariance-matrix BAO's can have a
several-percent effect on error-bar widths for some parameters
directly related to the BAO's, such as the baryon fraction.  Also, we
find that including the numerous galaxies in small haloes in a survey
can reduce error bars in these cosmological parameters more than the
simple reduction in shot noise might suggest.
\end{abstract}
\begin {keywords}
  cosmology: theory -- large-scale structure of Universe
\end {keywords}

\section{Introduction}
Humans are fortunate that there are baryons in the Universe.  Not only
are we made of them, but baryons are responsible for features in the
shape of cosmological power spectra and two-point correlation
functions that are quite valuable to cosmologists.  These features are
called baryon acoustic oscillations (BAO's), and are imprints of
acoustic oscillations in the gas of the early universe \citep{peeyu,
sz, holtz, ehu, mwp}.  Their presence in galaxy clustering statistics
provides a standard ruler to measure the relation between distance and
redshift, and the expansion of the universe at late times \citep{bg,
se}.  They provide one of the main tools currently proposed for
studying the effects of dark energy.

BAO's have been detected in modern low-redshift galaxy surveys, both
in the 2dFGRS \citep{c05} and SDSS \citep{e05,hutsi,pshape07}.  These
detections were made using the (two-point) correlation function, or
its Fourier dual, the power spectrum.  Conveniently, the BAO regime is
on large-enough scales that non-linear effects are mild.  Still, for
precision cosmology, these mild effects must be understood, and are
the topic of much recent research
\citep[e.g.][]{mill,jk,huff,sw,aea,se07,sssbao}.

Non-linear evolution of the matter power spectrum tends to dampen or
smear BAO's.  For example, \citet{esw} found that large-scale bulk
flows and cluster formation produce motions that smear out the BAO
peak in the correlation function, but that these motions are confined
to relatively small scales of $\sim 10 \hMpc$ in Lagrangian space.
They argued that these motions roughly preserve wiggles on the largest
scales of the power spectrum, but wipe out wiggles on smaller scales.
The effects we describe in this paper, using Eulerian perturbation
theory, likely arise physically from the same large-scale bulk motions.
The attenuation of BAO's can also be understood by considering an
additive mode-coupling power spectrum, which rises on small scales as
structure develops, along with a function which attenuates the linear
power spectrum on non-linear scales.  In the halo model \citep[HM,
reviewed in][]{cs}, this small-scale contribution is the one-halo (1h)
term, and comes from pairs of objects within single haloes.  A
qualitatively similar effect occurs in renormalized perturbation
theory \citep[RPT;][]{csrpt,mrpt}, which is less empirical than the
HM, and seems more accurate through translinear scales.  For example,
\citet{csrpt} show that, more physically than in the HM, the
mode-coupling power spectrum in RPT goes to zero on small scales.

In this paper, we show that wiggles exist in off-diagonal terms of
matter and galaxy power spectrum covariance matrices, almost entirely
out-of-phase with BAO's in the power spectrum.  We interpret these
BAO's in the covariance matrix as manifestations of the suppression
exacted on power-spectrum BAO's by power on large scales.  Regions of
the Universe with upward fluctuations in large-scale power have
more-suppressed BAO's.

The body of the paper is organized as follows.  In Section \ref{anal},
we discuss analytic predictions for the BAO's in the covariance matrix
of matter.  In Section \ref{nbody}, we test the analytic predictions
against $N$-body simulations, and investigate the detectability of the
wiggles in the covariance matrix.  Finally, in Section
\ref{cosmomatter}, we investigate the effect of these
covariance-matrix BAO's on cosmological parameter estimation, focusing
on parameters directly related to BAO's in the power spectrum.  We
perform this analysis for both matter and galaxy power spectra, using
the HM framework.

\section{Analytic predictions for covariance-matrix wiggles}
\label{anal}
The covariance of the matter power spectrum in a survey of volume $V$
(neglecting survey-shape effects) is the sum of a Gaussian term, which
depends on the square of the non-linear power spectrum, and a term
involving the matter trispectrum \citep[e.g.][HRS]{szh,hrs}.

\begin{equation}
C_{ij} = \frac{1}{V}\left[\frac{(2\pi)^3}{V_{s,i}}2P(k_i)^2\delta_{ij}
+ T_{ij}\right],
\label{cijdef}
\end{equation}
where $V_{s,i}$ is the volume of shell $i$ in Fourier space
(proportional to $k_i^3$ for logarithmically spaced bins), and
$T_{ij}$ is the parallelogram-configuration trispectrum averaged over
shells $i$ and $j$;
\begin{eqnarray}
  T_{ij} & \equiv & C(k_i,k_j)_{i\ne j}\nonumber\\
  & \equiv & \int_{s,i}\int_{s,j}T(\bk_i,-\bk_i,\bk_j,-\bk_j)\frac{d^3 \bk_i}{V_{s,i}}\frac{d^3\bk_j}{V_{s,j}}.
\label{angav}
\end{eqnarray}

\ptcovarwig

Figure \ref{ptcovarwig} shows the dimensionless, normalized
power-spectrum covariance $\tilde{T}(k_1,k) \equiv C(k_1,k)V/[P(k_1)^2
P(k)]$, for various $k_1$'s.  The wiggles in $\tilde{T}$ are much
stronger than in the power spectrum itself; the ratio of peak to
trough can be nearly a factor of ten.  However, as we discuss in
Sect.\ \ref{detect}, the amount of noise in a measurement of
$\tilde{T}$ is also large.

For Fig. \ref{ptcovarwig}, we use $P_{\rm lin}$ for the power spectrum
$P$, and the leading-order (third-order) PT trispectrum for $T_{ij}$.
In the upper panel, the curves are shown at redshift $z=0$; the curves
in the lower panel are independent of power-spectrum normalization and
redshift.  This is because the PT trispectrum involves three powers of
$P_{\rm lin}(k)$, evaluated at different $k$'s, as does the
denominator in the definition of $\tilde{T}$.  We use a linear power
spectrum from \camb\footnote{See \url{http://camb.info/}.}
\citep{camb}, with the same \LCDM\ cosmology as assumed for the
Millennium simulation \citep{mill}: the Hubble constant $H_0 = 73
\frac{\rm km}{\rm s~Mpc}$, $\Omega_{\rm CDM}=0.205$, $\Omega_b =
0.045$, the scalar spectral index $n_s = 1$, and the rms density
fluctuation in spheres of radius $8\ihMpc$ is $\sigma_8=0.9$.  The
dashed curve is an approximation for $k_1 \ll k$, given by
\begin{equation}
  \tilde{T}(k_1 \rightarrow 0, k) \approx \frac{5038}{2205} - 
  \frac{36}{35}\frac{P^\prime_{\rm lin}(k)k}{P_{\rm lin}(k)} + 
  \frac{1}{5}\frac{P^{\prime\prime}_{\rm lin}(k)k^2}{P_{\rm lin}(k)}.
\label{tapprox}
\end{equation}
The dotted curve shows this approximation without the first-derivative
term.  See Appendix \ref{ptwig} for a derivation of Eq.\
(\ref{tapprox}).

For $k_1 \ll k$, the dominant terms for a \LCDM\ power spectrum are
the second-derivative term, which produces wiggles $180^\circ$
out-of-phase with the wiggles in $P_{\rm lin}(k)$, and the constant.
The first-derivative term shifts the wiggles slightly.  The dominance
of the second-derivative term over the first-derivative term implies
that the main effect of large-scale power is that of suppressing
power-spectrum BAO's, not moving them.

By analogy with the non-linear power spectrum relative to the linear
power spectrum, it might seem that BAO's in the non-linear $\tilde{T}$
could be significantly attenuated on small scales compared to the PT
prediction.  However, leading order for the trispectrum is 3rd-order,
not 1st-order as for the power spectrum; thus, the PT trispectrum should be
valid into the mildly non-linear regime.  At $z=0$, 3rd-order PT works
to 1\% for the power spectrum for $k\lesssim 0.065\ihMpc$, while the
linear power spectrum fails at the $\sim1\%$ level already at
$k\approx0.01\ihMpc$ \citep{taka07}.

On small scales, the PT trispectrum prediction is expected to fail.
Probably the most plausible model currently on small scales is the
halo model, the matter trispectrum of which was worked out by
\citet[][CH]{chu}.

\hmcovarwig

Figure \ref{hmcovarwig} compares the HM and PT predictions
for $\tilde{T}$.  We continue to use $P_{\rm lin}$ in the denominator
of the definition $\tilde{T}(k_1,k) = C(k_1,k)V/[P(k_1)^2 P(k)]$, for
easy comparison of the pure PT prediction.  Alternatively, $\tilde{T}$
could be defined with the halo-model matter power spectrum in the
denominator.  For the HM prediction, we use the same implementation as
in \citet[][NS]{ns}; some details of the implementation also appear in Appendix
\ref{hmtri} below.  At $z=0$, the three-halo (3h) term contributes
significantly on large scales.  This cannot be the case at arbitrarily
small $k$, where PT should hold exactly.  This extra contribution in
the trispectrum over PT was noted by CH; it is similar to the
shot-noise-like one-halo term that contributes unphysical power on
large scales of the power spectrum \citep{cs,sea,csrpt}.  However, it could
still be that the 3h contribution is real at some mildly non-linear
scale, such as in the BAO regime.

\section{Testing against $\bmath{N}$-body simulations}
\label{nbody}
To test the analytic predictions, we ran a large suite of $N$-body
simulations to get a high signal-to-noise covariance matrix.  In
particular, we wanted to test our expectation that the added
covariance from the HM 3h term on large scales is artificial, and also
to test the detectability of the wiggles, since their amplitude seems
much larger than the wiggles in the power spectrum itself.

We ran 100 particle-mesh (PM) dark-matter simulations of box size 1024
$\hMpc$, each with 256$^3$ particles on a grid 256 cells on a side,
solved using the code of \citet{pmref}.  The cosmological parameters
we used are the same as used for the previous PT plots; for the
transfer functions, we used \camb.  For the initial conditions (IC's)
of the simulations, we used the 2LPT (2nd-order Lagrangian
Perturbation Theory) code \citep{s98}, run to $z=49$.  2LPT IC's have
reduced transients over IC's generated from the \citet{za}
approximation, and are also more accurate for higher-order statistics
\citep[see][and references therein]{cps}.  We also ran 400 2LPT
simulations to $z=0$, 100 of which had the same random-number seeds as
the PM simulations.

\showpows 

Figure \ref{showpows} shows average power spectra from the various
simulations.  We also show the \halofit\ non-linear power spectrum
\citep{sea}.  The average PM power spectrum departs clearly from
\halofit\ at $k \sim 0.14\ihMpc$, so this is roughly the point to
which we trust the PM simulations.  The second wiggle at $k\sim 0.12$
is attenuated in the PM power spectra relative to \halofit, but
\halofit\ was not developed with particular attention to BAO's, so the
PM simulations may still be trustworthy there.  The average 2LPT power
spectrum exhibits a gradual attenuation in power starting at $k\sim
0.05\ihMpc$, but we still analyse the 2LPT simulations at higher $k$
for their high-signal-to-noise covariance matrix, keeping in mind this
caveat.  Scoccimarro (private communication) speculates that the
deficit in power using 2LPT is from the large amount of shell-crossing
at $z=0$, and that smoothly truncating the initial power for $k\gtrsim
0.3\ihMpc$ would improve 2LPT's performance.

To measure the power spectra, we used FFT's run on a 256$^3$ mesh with
cloud-in-cell density assignment.  We corrected for the voxel window
function by extrapolating the factor between power spectra measured
from one of the simulations with 256$^3$ and 1024$^3$ grids.  We
subtracted off the Poisson shot noise of $64 \hMpcV$, even though it
is negligible over the range of scales we use.

Figure \ref{bigbins} shows measurements of $\tilde{T}(k)$ from a few
ensembles of simulations.  To measure $\tilde{T}$, we first measured
power spectra in each simulation, using large bins to reduce noise.
The bin edges were placed by hand to be approximately at the nodes
between expected wiggles in $\tilde{T}$.  The first bin runs from
$0<k<0.05 \ihMpc$; the lowest $k$ sampled was $2\pi/1024\hMpc \approx
0.006\ihMpc$.  The edge of the first bin, at $0.05 \ihMpc$, was chosen
to be about where the wiggles cease being prominent in Fig.\
\ref{ptcovarwig}.  When trying to detect the wiggles in $\tilde{T}$,
there is a trade-off in choosing the edge of the first bin; with
increasing $k$, the signal increases with the number of modes, but the
wiggles dampen.  It is possible that the wiggles could be more
prominent with another choice of bin edge.  Subsequent bin edges are
shown in the plot, where horizontal error bars end.  For each
ensemble, we calculated a covariance matrix from the simulations'
power spectra, and found $\tilde{T}(k)$ by dividing the covariance in
each bin by $P(k_1)^2P(k)$, where $P$ is the power spectrum measured
from the simulations, averaged in each bin and over all simulations.

\bigbins

The vertical error bars on the points at $k\sim 0.1\ihMpc$ were
obtained by jackknife resampling.  This method differs from
spatial-jackknife resampling which is often used in estimating errors
in large-scale structure; in spatial-jackknife resampling, errors are
estimated by excising small spatial regions from simulations.  In our
case, we form jackknife samples by excluding entire simulations
one-at-a-time from the sample.  The error bar width on an
$N$-simulation point is $\sqrt{N-1}$ times the standard deviation of
measurements of $\tilde{T}(k=0.1\ihMpc)$ from $N$ covariance matrices,
each formed by excluding one simulation at a time from the sample.

The outer points at $k\approx0.07$ and $0.12\ihMpc$ appear with cones
indicating the robustness of the peak at $k\sim 0.1\ihMpc$.  We do not
show these outer points with their own error bars because there is a
high degree of correlation among these three points.  Typically, the
curves for different jackknife subsamples just move up or down, and
the peak-to-trough contrast hardly changes.  The half-height of the
cone at each $k_i$ is the standard deviation (times $\sqrt{N-1}$) of
the differences $[\tilde{T}(0.1\ihMpc)-\tilde{T}(k_i)]$ measured from
the $N$ jackknife measurements.

The data in Figure \ref{bigbins} do not indicate any extra
contribution (such as the HM would give) to $\tilde{T}$ above the PT
prediction.  Both the PT and HM predictions are within the errors of
the two leftmost points for the PM ensemble, and these points may be
upward fluctuations in $\tilde{T}$ for the first 100 realizations.
For the 2LPT ensemble, adding 300 more realizations to the sample
moves the first two data points down, and the last point up.  So, we
recommend the simple PT covariance for the matter power spectrum for
$k\lesssim 0.1 \ihMpc$.  However, the PT covariance must fail at some
large $k$, so for analyses extending to larger $k$ than this, we would
recommend using the halo-model covariance, excluding the offending 3h
term.  This is rather arbitrary, but \citet[][NSR]{nsr} showed that
the 1h and (less so) the 2h terms dominate the contribution to Fisher
information loss in the halo-model matter power spectrum on non-linear
scales, and thus the 1h and 2h terms are important to include.  We do
not exclude the 4h term, since it is where the wiggles we are
investigating lie.

\fineweights 

Figure \ref{fineweights} shows $\tilde{T}$ measured with greater
resolution.  With these narrow bins, only for the 400-2LPT ensemble
was the level of noise low enough to view calmly.  The error bars are
the square roots of the diagonal elements of the $\tilde{T}$
covariance matrix, again jackknife-estimated.  To reduce the noise, we
also measured $\tilde{T}$ using the weightings method of HRS.  This
gives 52 estimates of the power spectrum per simulation, by passing
two differently phased long-wavelength sinusoidal window functions
through each vertex, face, and edge of the simulation box.  When
excluding a simulation from a jackknife sample using weightings, we
excluded all of its 52 power spectra at a time.

The reduction in noise from the weightings method is substantial, and
brings out the shape of $\tilde{T}$.  However, we noticed in this and
other investigations that the covariance estimated in this manner was
a bit higher than the covariance from the unweighted simulations.
This is likely a result of beat coupling to large scales that HRS
discuss.  Beat coupling refers to the extra covariance from
non-trivial geometry (roughly, edge effects) in a power spectrum
measured with finite bins.  The weightings, even though they are
minimally invasive, smear power among nearby wavenumbers as any survey
window function does.  When bins have a finite width, as they almost
always do, this smearing contributes to the covariance a trispectrum
term involving the power spectrum at the `beat' wavenumber (the
difference between the nearby wavenumbers).  We did not use the
weightings method for the investigations shown in Fig.\ \ref{bigbins},
but a method such as the weightings method would probably be necessary
to detect covariance-matrix wiggles in real surveys, and in doing
this, beat-coupling should be taken into account.  However,
beat-coupling seems approximately to cause a wholesale additive shift
in $\tilde{T}$, so its effects could be negligible if all one cares
about is the position of wiggles in $\tilde{T}$.  Because of beat
coupling, we view it as a coincidence that $\tilde{T}$ from the 400
weighted 2LPT simulations lies right on the PT estimate; its true
level for the 2LPT simulations is lower, as in Fig.\ \ref{bigbins}.
We suspect that the lower level of $\tilde{T}$ in the 2LPT ensemble is
related to 2LPT's diminishing accuracy on small scales, as shown by
the small-scale deficit in its power spectra, shown in Fig.\
\ref{showpows}.

\subsection{Detectability of covariance-matrix wiggles}
\label{detect}
We can also estimate roughly how big a survey is necessary to detect
BAO's in $\tilde{T}$ from Fig.\ \ref{bigbins}.  In 100 PM simulations
1024$\ihMpc$ on a side (with a total volume of $107 \hGpcV$), we
detected the peak at a 5-$\sigma$ (4.97-$\sigma$, to be exact) level.
This is the probability that $\tilde{T}(0.07) < \tilde{T}(0.1)$, as
estimated by dividing the mean of $[\tilde{T}(0.1)-\tilde{T}(0.07)]$
by the standard deviation of this quantity (the half-height of the
cone).  Doing the same on the right side of the peak gives an estimate
of 10$\sigma$ for the probability that $\tilde{T}(0.1) >
\tilde{T}(0.12)$.

We do not know precisely how the error bars in the difference between
the peak and the troughs shrink with volume, but we can roughly
estimate this by comparing the results from the ensembles of 100 and
400 2LPT simulations.  There are two separate effects reducing the
error bars: the increase in the number of power spectra used for the
covariance matrix, as well as the increase in raw volume.  The
relationship between error-bar width and volume is not necessarily a
simple power law.  Still, it provides an easy estimate: this
`difference error bar' at $k\approx0.7$ scales as $V^{0.22}$, and at
$k\approx 0.12$ it scales as $V^{0.35}$.  Taking the higher of these,
an estimate of the volume of a survey that would give a 1-$\sigma$
detection of the peak is $107/5^{1/0.35} \hGpcV = 1.1 \hGpcV$.
Similarly, 2, 3, and 4-$\sigma$ detections could be done with 8, 25,
and $57 \hGpcV$ surveys, respectively.  If we conservatively assume
that error bars $\propto \sqrt{V}$ (as one might first guess), then we
estimate that 1, 2, 3, and 4-$\sigma$ detections could be done with 4,
17, 39, and $68\hGpcV$ surveys.

\hmgalcovarwig

The wiggles in $\tilde{T}$ would almost certainly be measured with
galaxies, not matter.  Figure \ref{hmgalcovarwig} shows $\tilde{T}$ of
galaxies using the HM, using a few different halo occupation
distributions (HOD's).  These HOD's will be discussed in more detail
in Sect.\ \ref{cosmogal}, but for now, note that the wiggles in
$\tilde{T}$ for galaxies inhabiting smaller haloes are approximately
like those for the matter, and the wiggles dampen somewhat with
increasing host-halo mass.  However, some of this dampening is done by
the 3h term, which apparently contributes artificially to the HM
prediction in Fig.\ \ref{hmcovarwig}.  So, there is some reason to
doubt the dampening of the wiggles in $\tilde{T}$.  The situation is a
bit more complicated here than previously because we divide by $P_{\rm
g}$, the galaxy power spectrum for each galaxy sample, instead of the
linear power spectrum $P_{\rm lin}$, to get $\tilde{T}(k_1,k) =
C(k_1,k)V/[P(k_1)^2 P(k)]$.

Besides the differences in $\tilde{T}$ itself that galaxies would
introduce, several factors would complicate detecting $\tilde{T}$ in
reality.  Redshift space distortions and complicated survey geometries
would both need to be taken into account.  As the discrepancy with the
weightings method shows (Fig.\ \ref{fineweights}), even a minimal
window function affects measured covariances.  However, in that case,
it seems that the window function roughly adds a simple constant to
$\tilde{T}$, so it is possible that more complicated survey geometries
would also roughly preserve the shapes of the wiggles.

There are also factors that would tend to increase detectability.
First, as occurs exactly in PT, the galaxy $\tilde{T}$ is likely close
to independent of redshift, removing one of the usual complications of
measuring clustering statistics with a huge redshift survey.  Also, we
have not included the fact that a larger survey would allow
larger-scale modes $k_1$ to be accessed, where the wiggles in
$\tilde{T}(k_1,k)$ are more pronounced.  And, it is always possible
that there exists another method than we have used to detect the
wiggles more efficiently.

We have not explicitly investigated the issue of whether the wiggles
in $\tilde{T}$ could help significantly in fixing the baryon acoustic
scale, but it is worth further investigation, since every bit helps.
One might first guess that it would be easier to measure BAO's in a
three-point statistic than a four-point statistic such as $\tilde{T}$.
However, the particular parallelogram configurations of the
trispectrum which average to make $\tilde{T}$ depend only on the
amplitude of each Fourier mode.  Real numbers are not only easier to
deal with numerically, but we conjecture that statistics independent
of phases have smaller cosmic variance, since they are independent of
phase correlations which increase for smaller survey volume.  For
these reasons, and for the simplicity of the algorithm of the
estimator, these configurations of the trispectrum are competitive,
perhaps even advantageous over using the bispectrum for detecting
BAO's.  Thus it would be interesting to confirm our predictions in
data, since this would further corroborate our picture of
gravitational structure formation, and, in particular, demonstrate
that our understanding of BAO's extends to multipoint statistics.

\section{Effect on cosmological parameter estimation}
\label{cosmomatter}
Baryon oscillations have become quite popular as a prospective
cosmological probe.  Here we consider to what degree BAO's in the
covariance matrix affect these prospects, both for the matter power
spectrum and galaxy power spectra.  We investigate two parameters
that directly involve BAO's: the baryon acoustic scale, and the
baryon fraction $f_b\equiv \Omega_b/(\Omega_b+\Omega_{\rm CDM})$.  We
do not investigate the baryon acoustic scale directly; we investigate
the logarithm of the sound horizon when the baryons are released from
the Compton drag of photons \citep[see][EH]{ehu}.  This is
proportional to the baryon acoustic scale, i.e.\ the location of the
peak in the linear correlation function.

We use a Fisher matrix formalism \citep{fisher, tth} to estimate error
bars on these quantities, in the same way as in NS.  The cumulative
Fisher information in parameters $\alpha$ and $\beta$ over a range of
bin indices $i\in\R$ is approximated (assuming a Gaussian likelihood
function) as
\begin{equation}
  F_{\alpha\beta}(\R) =
  \sum_{i,j\in \R} \frac{\partial\ln P_i}{\partial\alpha}(\bssC_\R^{-1})_{ij} 
 \frac{\partial\ln P_j}{\partial\beta},
  \label{inforange}
\end{equation}
where $\bssC_\R$ is the square submatrix of $\bssC$ with both indices
ranging over $\R$.  For simplicity, in this paper, we just consider
single-parameter (unmarginalized) half-error bars; the half-error bar
in $\alpha$ is $\sigma(\alpha)\equiv 1/\sqrt{F_{\alpha\alpha}}$.

As recommended above in Sect.\ \ref{nbody}, we use the halo-model
trispectrum, excluding the 3h term, to compute the matter power
spectrum covariance matrix.  We do, however, show the results when the
full HM covariance (including the 3h term) is included; it does not
make a big difference.  We do not use simply the PT covariance because
we investigate well into the non-linear regime, up to almost $k=1
\ihMpc$.

\fbinfo

Figure \ref{fbinfo} shows what effect $T_{ij}$ terms in the covariance
matrix have on estimation of the baryon fraction from the matter power
spectrum.  For this and subsequent figures, we assume a fixed volume
of $1 \hGpcV$.  Each curve starts on the right at a $k_{\rm max}$, and
shows how error bars in $\ln f_b$ tighten as the power spectrum is
measured over an increasing range of $k$.  Four curves depart from
each $k_{\rm max}$ curve, calculated using different covariance
matrices: a purely Gaussian covariance matrix, including just the
$P(k)^2$ term on the diagonal; a `HM-3h' covariance matrix, our
fiducial covariance matrix, which uses the full HM except for
the 3-halo term; a full halo-model covariance matrix; and a `HM-3h,
nowig' covariance matrix, which uses a no-wiggle transfer function of
EH instead of \camb\ to obtain the input power spectrum.  We normalize
the `nowig' power spectrum to match the \camb\ power spectrum in
amplitude at $k=10^{-5} \ihMpc$, and also set it equal to the \camb\
power spectrum for $k>130 \ihMpc$, the wavenumber where the two power
spectra cross on scales smaller than the BAO regime.  Fig.\ \ref{hods}
(mainly about galaxy power spectra, discussed in Sect.\
\ref{cosmogal}) shows the halo-model matter power spectrum, along with
its no-wiggle counterpart.

Going well into the non-linear regime (looking at the blue curve, with
$k_{\rm max} = 0.6 \ihMpc$), the wiggly covariance matrix gives about
7\% tighter error bars, going to small $k$, than the non-wiggly
covariance matrix.  The difference is also 7\% if a full HM covariance
matrix is used for each.  The purely Gaussian covariance matrix
clearly gives underestimated error bars here.  This is because of the
`translinear information plateau;' significant correlations arise from
halo mass-function fluctuations between power-spectrum bins in the
range $0.2\lesssim k/(\nospaceihMpc) \lesssim 0.8$ \citep[][NSR,
NS]{rh05, rh06}.

However, in an intermediate regime (looking at the green curve, with
$k_{\rm max} = 0.2 \ihMpc$), the full covariance matrix actually gives
tighter error bars than if just the Gaussian term is used.  Here, the
wiggly covariance matrix gives tighter error bars by 4\% than the
non-wiggly one.  Starting on larger scales (looking at the red curve,
with $k_{\rm max} = 0.06 \ihMpc$), the covariance matrices including
$T_{ij}$ terms again give somewhat larger error bars than if only the
Gaussian term is used.

The derivative terms $\partial\ln P(k)/\partial\ln f_b$ are the same
for all calculations in Fig.\ \ref{fbinfo}.  For this, we calculated linear
power spectra with \camb\ using slightly different baryon fractions,
at fixed $\sigma_8$.  We then put these varied linear power spectra
through our halo-model code.  The rise of the one-halo term on small
scales does attenuate the BAO's somewhat compared to the linear power
spectrum, but not to the degree seen in $N$-body simulations.  We thus
used the method described by \citet[][P07]{p07}, based on the work of
\citet*{esw}.  The wiggles are attenuated by Gaussian window in Fourier
space of width $1/(10\hMpc)$.

\shinfo

Figure \ref{shinfo} is the same as Fig.\ \ref{fbinfo}, except it shows
error-bar tightening in a quantity more popular in BAO parameter
estimation, the baryon acoustic scale (we actually show error bars in
the logarithm of the sound horizon).  In this case, the error bars
using the wiggly covariance matrix differ nearly indistinguishably
from those using the non-wiggly ones; going from arbitrarily large
scales to $k_{\rm max}=0.3$, there is only a 2\% difference in the
error bars.  In fact, excluding all $T_{ij}$ terms makes at most a 6\%
difference in the error bars.  The `Full HM' curves have been left off
this plot, since they are indistinguishable from the `HM-3h' curves.

In Fig. \ref{shinfo}, again we use the same derivative terms $\partial
\ln P(k)/\partial \ln s$ for each curve.  We estimate $\partial \ln
P(k)/\partial \ln s$ by changing the sound horizon by hand in the
transfer function code of EH.  This is not terribly physical, but it
does what we want: changing $s$ by hand in the EH code results in
power spectra which are effectively identical, except that the BAO's
are offset.  As with our investigation of the baryon fraction, we
attenuate the baryon wiggles with a Gaussian window.

It is interesting that the wiggles in the covariance matrix generally
result in somewhat narrower constraints in $\ln f_b$.  Intuitively, it
makes sense that knowledge of how BAO's are attenuated by fluctuations
in large-scale power would help to constrain $f_b$, which directly
affects the BAO amplitude.

The effect of $\tilde{T}$ is smaller for sound-horizon determination.
Numerically, this is for two reasons: one, the derivative term
$\partial \ln P(k)/\partial \alpha$ is effectively zero on large
scales.  Also, the wiggles in the derivative term are well
out-of-phase with wiggles in the Fisher matrix, $\bssC^{-1}$.  The
difference might be greater if the BAO's in $\tilde{T}$ were not so
close to entirely out-of-phase with the BAO's in the power spectrum;
that is, $\tilde{T}$ might have a greater effect if large-scale power
appreciably moved the BAO's.

\subsection{Wiggles in galaxy power spectra}
\label{cosmogal}

In the previous section, we showed that BAO's in the covariance matrix
of the matter power spectrum have a several-percent effect on the
inferred error bars on the baryon fraction $f_b$, and a smaller
effect on error bars in the sound horizon $s$.  Here we investigate to
what degree these effects propagate through to galaxy power spectra.

We use a halo occupation distribution \citep[HOD, e.g.\ ][]{bw} to
model the galaxy power spectrum and covariance matrix.  We build off
of the matter trispectrum as worked out by CH, and use the satellite
HOD as introduced by \citet[][K04]{k04}, with its simple Poisson
satellite HOD moments.  See Appendix \ref{hmtri} for details of our
implementation.

In the K04 model, a halo has a central galaxy if and only if the halo
has a dark-matter mass $m > M_{\rm min}$.  The other two parameters in
the HOD are $M_1$, the mass at which, on average, a halo has one
satellite galaxy, and $\gamma$, the slope in the formula giving the
mean number of satellites in a halo as a function of mass;
\begin{equation}
\langle N_s|m\rangle = (m/M_1)^\gamma.
\label{k04ns}
\end{equation}

Three parameters give a huge potential parameter space to explore.  To
narrow it, we vary $M_{\rm min}$, and fix $M_1/M_{\rm min} = 30$
(which K04 suggested as a fiducial value).  Observed galaxy power
spectra at $z=0$ are generally close to power laws, so to use the most
physically plausible HOD's, we set $\gamma$ at each $M_{\rm min}$ by
maximizing the galaxy power spectrum's straightness.  More precisely,
we minimized the sum of the squares of the second derivative of
$P_{\rm g}(k)$ in log-log space over a range
$10^{-0.5}<k/(\nospaceihMpc)<100$.  The three halo masses $10^{11.5}$,
$10^{12.5}$, and $10^{13.5} M_\odot$ have best-fitting values
$\gamma=0.97$, $1.23$ and $1.68$.

\hods

To account for galaxy shot noise, we use the shot-noise-added power
spectrum $(P_{\rm g}(k) + 1/\bar{n}_{\rm g})$ for the Gaussian terms
on the diagonal of the covariance matrix \citep[e.g.][]{c04}.  Here,
$\bar{n}_{\rm g}$ is the number density of galaxies.  This treatment
of shot noise is equivalent to using an `effective volume' $V^{\rm
eff}(k) = V/\{1+1/[\bar{n}_{\rm g}P_{\rm g}(k)]\}^2$ at each $k$
\citep[e.g.][]{teg}.  The shot noises for the three HOD's used, in
increasing order of $M_{\rm min}$, are $1/\bar{n}=6200$, 540, and
$60\ihMpcV$.  For the smallest-$M_{\rm min}$ sample, the shot noise is
negligible up to $k = 1 \ihMpc$; for the two larger ones, the shot
noise equals $P_{\rm g}$ at $k\approx 0.9$ and $0.2 \ihMpc$.

Figure \ref{hods} shows the power spectra for the HOD's we investigate
below, along with the matter power spectrum investigated in Sect.\
\ref{cosmomatter}, the \camb\ linear power spectrum, and the \halofit\
non-linear power spectrum.  We also show each power spectrum as
produced from the EH no-wiggle power spectrum.  Each power spectrum is
divided by the EH no-wiggle linear power spectrum for clarity.  The
wiggles in the matter and galaxy power spectra here are only
attenuated by the rising, unwiggly 1h term.  The only place where we
implement the more-accurate method of P07 is in the derivatives of the
power spectrum with respect to parameters, which is where it matters
significantly for parameter estimation.  We use the same P07 BAO
attenuation for all galaxy and matter power spectra, since the
attenuation in this model is a function of a single parameter, the
width of a Gaussian window.  It is unclear how to change this
parameter for different galaxy populations.

\galfbinfo
\galshinfo

Figures \ref{galfbinfo} and \ref{galshinfo} show error-bar tightening
in the baryon fraction and the sound horizon, in a similar manner as
in Figs.\ \ref{fbinfo} and \ref{shinfo}.  Here, we fix $k_{\rm
max}=0.3\ihMpc$, but show results for three different $M_{\rm min}$'s.
We include the 3h term in the galaxy covariance matrix since we do not
have theoretical arguments and measurements (as we do for matter)
telling us to discard it.

At our fixed volume $V = 1 \hGpcV$, constraints are tighter for
galaxies with a lower $M_{\rm min}$.  There are two causes for this.
First, galaxies with high $M_{\rm min}$ are sparser and therefore have
more shot noise, an effect which wins out over their increase in bias.
The shot-noise effect can roughly be judged by looking at the dotted
curves at different $M_{\rm min}$'s.  The difference between the
brightest and intermediate HOD's is mainly from this.

Second, the dominance of lower-halo (1h, 2h, and 3h) terms grows with
$M_{\rm min}$ in the HM covariance matrix.  This is what causes the
drastic difference between the middle and faintest samples.  It also
increases the impact of covariance-matrix wiggles for the faintest
galaxies; for this sample, the difference in error-bar width between
the wiggly and non-wiggly covariance matrices reaches 13\% for $\ln
f_b$, and 4\% for $\ln s$.  The lower-halo terms attenuate the wiggles
in $\tilde{T}$ increasingly for brighter galaxy samples, as shown in
Fig.\ \ref{hmgalcovarwig}.

As $M_{\rm min}$ increases, not only are the effects of lower-halo
terms greater on large scales, but the scale increases (i.e.\ the
critical $k$ reduces) where lower-halo terms come to entirely dominate
the covariance.  Thus, if $M_{\rm min}$ is small, the wiggles in the
covariance matrix can be accessed over a larger range of scales.
Roughly, if the 1h term comes in on smaller scales in the power
spectrum (as it does for the fainter galaxy samples), then more
cosmological information can be measured from the 2h (quasi-linear)
term.

This is similar to what we found in NS; cosmological information is
more pristine in regions of the universe in earlier stages of
structure formation.  It seems likely that, again as in NS, imposing
an $M_{\rm max}$, i.e.\ excluding galaxies in large clusters, could
help significantly to tap smaller-scale information.  However, just as
in that study, we find that our model breaks down when the lowest-mass
haloes dominate the power spectrum and its covariance matrix.  For
$M_{\min} \lesssim 10^{11} M_\odot$, using values of $\gamma$ fit in
the same way as for the samples we display, we sometimes get
non-positive-definite covariance matrices.  (However, it is possible
with slightly different $\gamma$'s to get well-behaved covariance
matrices.)  

We attribute this breakdown for small $M_{\rm min}$ to our lack of
accurate knowledge about the power spectrum and trispectrum of haloes
($P_{\rm hh}$ and $T_{\rm hhhh}$) in the translinear regime.
\citet{sss} showed that $P_{\rm hh}$ is difficult to model; $T_{\rm
hhhh}$ is likely harder.  Our assumption that leading-order PT
describes the polyspectra of haloes is not unreasonable, but the
leading order is different for each polyspectrum.  Using different
orders of PT for the power spectrum and trispectrum together is
inconsistent in the translinear regime, where higher-order corrections
to the power spectrum are significant.  For the matter power spectrum,
and for high $M_{\rm min}$, the 1h terms of both the power spectrum,
and its covariance, dominate terms involving $P_{\rm hh}$ and $T_{\rm
hhhh}$ on translinear scales.  In these cases, accurate modelling of
$P_{\rm hh}$ and $T_{\rm hhhh}$ would make a negligible difference in
our calculations.  When the 1h terms are reduced relative to other
terms, though, the raw $P_{\rm hh}$ and $T_{\rm hhhh}$ are exposed in
a regime where they are not known accurately.

Because the $M_{\min} = 10^{11.5} M_\odot$ sample is close to the
critical $M_{\min} = 10^{11} M_\odot$ where covariance matrices can be
non-positive-definite for fiducial $\gamma$'s and $M_1$'s, we view the
drastic diminution in the error bars using this sample with some
caution.  Still, we confidently assert that in the HM, reducing
$M_{\rm min}$ can produce significantly smaller error bars than the
simple reduction in shot noise would suggest.

Even though there seem to be significant gains in cosmological
information when using a fainter galaxy sample, it is important to
remember that we are holding the volume of the survey fixed.  We have
not considered some important realistic effects, an obvious one being
that it takes less telescope time to measure bright than faint
galaxies.

The question of how to choose a galaxy sample to maximize cosmological
information is a very important and interesting one, but it lies
outside the scope of this paper.  The answer to the main question of
this section is this: wiggles in galaxy power-spectrum covariance
matrices do seem to affect constraints on $\ln f_b$ and $\ln s$ to
about the same (if not somewhat larger) degree as for matter.

\section{Conclusions} 
Our main points are the following:

\begin{itemize}
\item In off-diagonal entries in power spectrum covariance matrices,
BAO's exist which appear much stronger than the BAO's in the power
spectrum.  These wiggles are a manifestation of the suppression which
large-scale power does to BAO's in the power spectrum, and originate
in the perturbation-theory trispectrum.  We give a simple analytic
approximation to these wiggles in terms of the linear power spectrum
and its first two derivatives, and check the analytical predictions
using $N$-body simulations.

\item These wiggles are potentially detectable in current and upcoming
surveys, but because of the large noise in a covariance matrix
measurement, they are only detectable at a small significance level.
We estimate that a one-sigma detection could be done with a survey of
size a couple of $h^{-3}{\,}{\rm Gpc}^3$, and that a three-sigma
detection could require a survey of volume $\sim30\hGpcV$.

\item The wiggles make a modest difference in cosmological parameter
error bars from analysing galaxy and matter power spectra.  For
example, using the true, wiggly covariance matrix in estimating the
baryon fraction results in error bars several percent tighter than a
no-wiggle covariance matrix.  Doing so in estimating the baryon
acoustic scale has a smaller effect.

\item At fixed volume, including galaxies in smaller-mass haloes
provides tighter error bars on parameters such as the baryon fraction
and the baryon acoustic scale than analysing galaxies in only large
haloes.  In the context of the HM, this effect goes beyond the
simple gains from analysing a sample with smaller shot noise.  We
attribute this to the one-halo term of the galaxy power spectrum
becoming dominant at smaller scales for small haloes than large ones.
\end{itemize}

\begin{sloppypar}
Most of the calculations in this paper made use of our package of Python
code for cosmology, called {\scshape CosmoPy}. It can be downloaded
from \url{http://www.ifa.hawaii.edu/cosmopy/}.
\end{sloppypar}

\section*{Acknowledgments}
We thank Adrian Pope, Ben Granett, Martin White, Alex Szalay and Tom
Bethell for helpful discussions, Nick Gnedin for access to his PM
code, and an anonymous referee for perspicacious questions and
comments.  We are grateful for support from NASA grant NNG06GE71G, and
NSF grants AST-0206243, ITR 1120201-128440, and AMS04-0434413.

\onecolumn
\appendix
\section{Wiggles in the perturbation-theory trispectrum}
\label{ptwig}

The expression for the 3rd-order PT trispectrum includes 12 terms
involving the $F_2$ symmetrized kernel, and 4 terms involving the
$F_3$ symmetrized kernel.  In this Appendix, all appearances of $F_2$
and $F_3$ are the symmetrized kernels, often denoted in the literature
with a superscript $^{(s)}$.  Expressions for these mode-coupling
kernels can be found in, for example, \citet{goroff} and \citet{bea}.
The $F_2$ kernel is
\begin{equation}
F_2(\bk_1,\bk_2)=\frac{1}{2}\left[1+\mu+\left(\frac{k_1}{k_2}+\frac{k_2}{k_1}\right)c + (1-\mu)c^2\right],
\end{equation}
where $c\equiv \cos\theta_{12}$.  In the following, we take the
Einstein-de Sitter value of $\mu=3/7$, even though it has a mild
dependence on $\Omega_m$; $\mu\approx\frac{3}{7}\Omega_m^{-2/63}$
\citep{bjcp}.  This quantity $\mu$ also appears in the $F_3$ kernel.
For calculations, we do use this expression for $\mu$; for our assumed
cosmology, $\mu\approx 1.045\times\frac{3}{7}$.  The PT trispectrum
is

\begin{equation}
  T(\bk_1,\bk_2,\bk_3,\bk_4) = 4[F_2(\bk_{13},-\bk_3)F_2(\bk_{24},-\bk_4)P_3 P_4 P_{13} + {\rm cyc.}] + 6[F_3(\bk_2,\bk_3,\bk_4)P_2 P_3 P_4 + {\rm cyc.}].
\label{tdef}
\end{equation}
Here, $k_{13} = k_1+k_3$ and $P_{13} = P(k_{13})$; in this Appendix,
$P$ denotes $P_{\rm lin}$, the linear power spectrum.

The abbreviation `cyc.' describing the 12 $F_2$ terms could do with
some further explanation.  The $F_2$ terms come from terms such as
$\langle\delta^{(2)}(\bk_1)\delta^{(2)}(\bk_2)\delta^{(1)}(\bk_3)\delta^{(1)}(\bk_4)\rangle_{\rm
c}$, where the superscript on $\delta$ is the order of perturbation.
There are six choices of where to put the $^{(2)}$ labels.  Each of
these gives two $F_2$ terms;
\begin{equation}
  \langle\delta^{(2)}(\bk_1)\delta^{(2)}(\bk_2)\delta^{(1)}(\bk_3)\delta^{(1)}(\bk_4)\rangle_{\rm c} = F_2(\bk_{13},-\bk_3)F_2(\bk_{24},-\bk_4)P_3 P_4 P_{13} + F_2(\bk_{14},-\bk_4)F_2(\bk_{23},-\bk_3)P_3 P_4 P_{14}.
\label{explic}
\end{equation}

The PT trispectrum contribution to the power-spectrum covariance can
be simplified to the following, given by \citet{szh} (which we
rearrange): 
\begin{eqnarray} T_{ij} & = & \int_{k_i}\int_{k_j}
\left\{12 P_1 P_2 \left[F_3(\bk_1,-\bk_1,\bk_2)P_1 +
F_3(\bk_2,-\bk_2,\bk_1)P_2\right] +
8P_{1-2}\left[F_2(\bk_{2-1},\bk_1)P_1 +
F_2(\bk_{1-2},\bk_2)P_2\right]^2
\right\}\frac{d^3\bk_1}{V_s(k_i)}\frac{d^3\bk_2}{V_s(k_j)}.\nonumber\\
\label{rearrszh} \end{eqnarray} Here, $k_{1-2} = k_1-k_2$, and
$P_{1-2}=P(|k_{1-2}|)$.

A possibly useful intermediate result for the $F_2$ part is, again where $c=\cos\theta_{12}$, 
\begin{equation}
F_2(\bk_{2-1},\bk_1)P_1 + F_2(\bk_{1-2},\bk_2)P_2 = \frac{\left[(3-10c^2)k_1+7ck_2\right]k_2^3P_1 + (k_1 \leftrightarrow k_2)}{14k_1k_2(k_1^2+k_2^2-2ck_1k_2)}.
\end{equation}
Where this time $c=\cos\theta_{ab}$, the $F_3$ part can be expressed as \citep{val}:
\begin{equation}
F_3(\ba,-\ba,\bb)=\frac{b^2\left[a^4(10-59c^2+28c^4)+a^2b^2(10-44c^2+76c^4)-21b^4c^2\right]}{126a^2\left[a^4+b^4+2a^2b^2(1-2c^2)\right]}.
\end{equation}

The main approximation we use is $k_i \ll k_j$; we keep terms up to
zeroth order in $k_1$ (terms in $k_2/k_1$ to the powers 2, 1, and 0).  Substituting
$P(k_2)-(ck_1)P^\prime(k_2)+\frac{1}{2}(ck_1)^2P^{\prime\prime}(k_2)$
for $P_{1-2}$ and averaging over $c$ gives
\begin{equation}
T_{ij} = \left[\frac{5038}{2205}P(k_j) - \frac{36}{35}P^{\prime}(k_j)k_j + 
\frac{1}{5}P^{\prime\prime}(k_j)k_j^2\right]P(k_i)^2,
\label{tijapprox}
\end{equation}
which is the result in Eq.\ (\ref{tapprox}).
There is no term involving a higher derivative of $P(k)$ to this order in
$k_i$.  Another approximation which makes a term involving
$P(k_i) P(k_j)^2$ disappear in Eq. (\ref{tijapprox}) is $\frac{k_i^2}{P(k_i)} \ll
\frac{k_j^2}{P(k_j)}$.

\section{The halo-model galaxy trispectrum}
\label{hmtri}

The galaxy trispectrum, excluding the effects of shot noise, has a
similar form to the matter trispectrum, which CH worked out.  The main
difference is that matter densities get replaced by galaxy
number densities.

In the HM, galaxies and dark matter are assumed to lie
entirely within haloes.  Like CH, we assume that the haloes are
distributed according to leading-order perturbation theory (PT), i.e.\
the linear power spectrum, and the 3rd-order PT trispectrum.  This
is not quite true \citep{sss}, but there seems currently to be no
better analytic approximation to use.

The matter power spectrum in the HM is the sum of two terms:
\begin{equation}
P_{\rm mm}(k) = P^{\rm 1h}_{\rm mm}(k) + P^{\rm 2h}_{\rm mm}(k) = M^0_2(k,k) + P^{\rm lin}(k)[M^1_1(k)]^2,
\label{p1h2h}
\end{equation}
where $M^\beta_\alpha$ are integrals over the halo mass function $n(m)$.
\begin{equation}
M_\alpha^\beta(k_1,\ldots,k_\alpha) \equiv \int_0^\infty
(m/\bar{\rho})^\alpha u(k_1,m)\cdots u(k_\alpha,m) b_\beta(m)n(m)\,dm.
\label{malphabeta}
\end{equation}
Here, $\bar\rho$ is the mean matter density, $b_\beta(m)$ is the
$\beta$-order halo bias \citep[][SSHJ]{mjw,sshj}, and $u(k,m)$ is the
Fourier-transformed halo matter-density profile, for which we take a
\citet{nfw96} form.  In principle, $u$ could be complex, but we assume
a spherically symmetric form, which forces it to be real.  We use a
\citet{st} form for the halo mass function.  We also integrate over a
\citet{bull} concentration distribution at each mass, but suppress
that integral in these equations for simplicity.

Our implementation of the galaxy trispectrum uses the subhalo HOD of
K04.  There are other parameterizations of the HOD, for example a
binomial model by SSHJ, but the K04 model gives a simple form for
moments of the HOD.  The parameters of the K04 model are explained in
Eq.\ (\ref{k04ns}), and the text above it.

The number of satellites is assumed to obey a Poisson distribution.
This gives a simple formula for its factorial moments, which
\citet{k04} found to hold through the third moment.  Where $N_s^{(\alpha)}$
is the $\alpha$th factorial moment of the satellite HOD,
\begin{equation}
N_s^{(\alpha)}(m) \equiv \langle N_s(N_s-1)...(N_s-(\alpha-1))|m\rangle = \langle N_s|m\rangle^\alpha = (m/M_1)^{\alpha\gamma}
\end{equation}

The dark matter trispectrum has 1h, 2h, 3h, and 4h terms, for which we
refer the reader to CH.  When converting these to galaxy trispectrum
terms, the $M^\beta_\alpha$ factors are replaced with
$G^\beta_\alpha$, defined as
\begin{equation}
G_\alpha^\beta(k_1,\ldots,k_\alpha) \equiv \int_{M_{\rm min}}^\infty \ug(k_1,m)\cdots \ug(k_\alpha,m) \frac{1}{\bar{n}_{\rm g}^\alpha}\left[N_s^{(\alpha)}(m) + N_s^{(\alpha-1)}(m)\sum_{i=1}^{\alpha}\frac{1}{\ug(k_i,m)}\right]b_\beta(m)n(m)\,dm.
\label{galphabeta}
\end{equation}
This is a simplification of expressions by \citet[][e.g.\ their Eq.\
76]{sws} for the case of the satellite HOD.  Here, $\bar{n}_{\rm g}$
is the mean galaxy number density.  For simplicity, we assume that the
halo galaxy-number-density profile $\ug$ is real, since we deal
exclusively with spherically symmetric haloes.  Ignoring the second
term in brackets and the lower limit of integration, this expression
has a nearly identical form to $M_\alpha^\beta$ in Eq.\
(\ref{malphabeta}).  The second term in brackets shows the
contribution by multiplets of galaxies with the central galaxy.  The
central galaxy is assumed to lie in the centre of its halo, so in real
space, the term including it gets one fewer convolution over the
galaxy-density profile.

For example, the galaxy power spectrum is 
\begin{eqnarray}
P_{\rm g}(k) & = & P^{\rm 1h}_{\rm g}(k) + P^{\rm 2h}_{\rm g}(k) = G^0_2(k,k) + P^{\rm lin}(k)[G^1_1(k)]^2\nonumber\\
& = & \int_{M_{\rm min}}^\infty \frac{N_s^{(2)}(m)\ug(k,m)^2 + 2 N_s^{(1)}(m) \ug(k,m)}{\bar{n}_{\rm gal}^2} n(m)\,dm +
P^{\rm lin}(k) \left[\int_{\rm min}^\infty \frac{N_s^{(1)}(m)\ug(k,m) + 1}{\bar{n}_{\rm gal}}b_1(m)n(m)dm\right]^2.
\label{pgal}
\end{eqnarray}
We note that expressions for the 2h part of the halo-model galaxy
power spectrum in the literature sometimes omit the second term in the
2h numerator (the number $1$, which comes from pairs of central
galaxies with each other), even though it dominates the first term in
the 2h numerator on small scales.  However, the 1h term is dominant
over the 2h term on small scales, so the central-galaxy contribution
to the 2h term usually contributes negligibly to the overall $P_{\rm
g}$.

For the galaxy power spectrum and trispectrum, we do not change the
galaxy-density halo profiles or concentrations from what we used for
the dark matter.  As with the matter covariance matrix, we integrate
over concentration parameter at each mass.  Galaxy-density profiles
are almost certainly different from matter-density profiles in reality
\citep[e.g.][]{nk}, but we are unaware of a simple analytic
alternative to what we used for the dark matter.  In any case, these
approximations are adequate for our present purposes.

\end{document}